\documentclass{article}
\usepackage{graphicx}
\usepackage{epsfig}

\title{An Optimal Lookup Table Construction for Charge
Division with respect to Efficiency Flatness.}

\begin{document}

\maketitle
\begin{center}
\author{P. Van Esch \footnote{Corresponding author} \\ \and F.
Millier

 \textit{Institut Laue Langevin \\ 6, rue Jules Horowitz \\ BP 156 \\
F-38042 Grenoble \\ France}}
\end{center}

\begin{abstract}
Static lookup tables as often used in the position calculation
electronics in position sensitive detectors suffer from the well
known problem that the propagation of digitization errors in the
division leads to unequal efficiencies for the different output
channels. In this paper we present a solution to the construction
of such a lookup table which is optimal in the sense that the
least possible efficiency errors are committed in the class of
monotone lookup tables. The obtained resolution is comparable to
what is obtained using the standard implementation of the fixed
point division. After a theoretical explanation, we present some
experimental results confirming our claim.
\end{abstract}

\pagestyle{myheadings}
\markright{Optimal Lookup Table in Charge Division}
\section{Introduction}

Resistive charge division is a well-known technique for
determining the position of the impact of an ionizing particle
along a resistive electrode read out by charge amplifiers on both
sides.   If X is the charge measured at side A and Y is the charge
at side B, then the position can ideally be calculated from the
dimensionless number:
\begin{equation}
P = \frac{X}{X+Y}
\end{equation}
Several electronic techniques exist to calculate the relative
position P from the electronic signals X and Y. Although analog
treatment has been used in the past with success, the progress in
ADC converters and digital circuitry promotes an all-digital
solution.   Moreover, the existence of large EPROM memories
invites the use of an extremely simple design, which is often used
in correspondence with position sensitive detectors: a lookup
table which associates which each couple \( (X,Y) \) the position
\(P\). In practice, we've build a circuit with two 11-bit inputs
for \(X\) and \(Y\) and one 8-bit output for \(P\).

\section{Digitization Problem}

Although the task of filling in the lookup table may seem simple
(use the digitized values of X and Y, and calculate P, rounded to
8-bit accuracy) it leads to an unequal efficiency of each of the
256 channels due to digitization effects.  Perfectly uniform
irradiation would then lead to unequal intensities observed in the
different channels. In the rest of this paper, we will refer to
this assignment as the 'standard position calculation'. The
quantization effect and the reasons why they are sometimes a
problem are well-described in \cite{articlegeesman}. In this paper
we set out to find the best possible lookup table which reduces
this effect maximally, without compromising the resolution.

The effect can be explained as follows: a true value X will be
represented by its digitized counterpart \(X_{d}\) and a true
value Y will be represented by \(Y_{d}\).  So a digital couple
\((X_{d},Y_{d})\) represents a square in the \((X,Y)\) plane with
center of gravity \((X_{d}+1/2,Y_{d}+1/2)\). The standard way of
assigning this point (and so all couples in the square we're
considering) to a position bin is by calculating:
\begin{equation}\label{eq:continuousdiscrete}
    P'= \frac{X_{d}+1/2}{X_{d}+Y_{d}+1}
\end{equation}
and after digitization:
\begin{equation}
    P_{d}= [\frac{X_{d}+1/2}{X_{d}+Y_{d}+1}2^{N_{out}}]
\end{equation}
Here, \(N_{out}\) is the number of bits in the output (for 256
position channels, it is equal to 8). The problem arises because
this assignment maps the whole square of \((X,Y)\) values to the
value \(P_{d}\) as determined by the exact position of its center,
and the distribution of these center points is not uniform over
the 256 position channels. The probability for a point \((X,Y)\)
to be solicited depends on \((X + Y)\) through the pulse height
spectrum of the incident radiation.

We make one fundamental hypothesis: we assume that the pulse
height spectrum of the captured radiation is independent of
position.  So we're actually considering an event distribution
which factorizes in the variables E and P:
\begin{equation}
d\sigma = \textrm{spec}(E) \textrm{image}(P) dP dE
\end{equation}
We know \(\textrm{spec}(E)\) (the pulse height spectrum, assumed
independent of the position), and we want to find out
\(\textrm{image}(P)\) from \(\sigma(E,P)\). However, technically,
we're working in the variables X and Y, so we have:
\begin{equation}
d\sigma = \rho(X,Y)dX dY
\end{equation}
So our task is to extract \(\textrm{image}(P)\) from
\(\rho(X,Y)\). Working directly in the digitized variables (but
analytically continuing them...), we have the following coordinate
transformation:
\begin{eqnarray}
\nonumber
  E &=& X + Y + 1 \\
  P &=& \frac{X + 1/2}{X + Y + 1}
\end{eqnarray}
The Jacobean of the transformation \(\partial(P,E)/\partial(X,Y) =
1/(X+Y+1) = 1/E\), so we obtain:
\begin{equation}
    \rho(X,Y) = \frac{\textrm{spec}(E)}{E}\textrm{image}(P)
\end{equation}
This suggests that points which are geometrically uniformly
distributed in the \( X,Y \) plane (as our digitized
\((X_{d},Y_{d}) \) couples are)  have a weight equal to
\(\textrm{spec}(E)/E \) in the image. It would hence be logical
that each bin of the image (each digitized slice of the variable
P) contains as much as possible an equal total weight of \(
\textrm{spec}(E)/E \) over each of the points \((X_{d},Y_{d})\)
assigned to the bin in the lookup table. Of course there is the
constraint that the P'-value of each couple should still be close
to the nominal P-value of the chosen bin (otherwise the resolution
of our image will suffer seriously).  We propose a method which
satisfies both requirements.

\section{Proposed method}

There are already several successful techniques that are proposed
to tackle this problem: \cite{articletakahashi} proposes to assign
the (X,Y) point partly to different P-bins ; \cite{articlekoike}
extends the effective word length over which the division is
worked out. \cite{articleberliner} partly uses a lookup table to
invert (X+Y) and uses a processor to calculate the final position.
However, all these techniques change the basic electronic
architecture and do not use a simple lookup table.
\cite{articlemori} and \cite{articleuritani} have written about
half-analog methods, using logarithmic amplifiers.
\cite{articleuritani} has moreover proposed a more classical
method of correcting the image off line with efficiency correction
coefficients which can be calculated from the pulse height
distribution.  He points out that this gives rise to increased
statistical errors.  \cite{articlegeesman} does give an improved
way of calculating a lookup table, but the method is sub-optimal,
although probably sufficient for many applications.

We set out to find a mathematically optimal solution of
constructing a lookup table.  An optimal solution can eventually
be used as a benchmark to compare to other techniques.

We first create a list L of all possible pairs \((X_{d}, Y_{d})\)
of which there are, in our case, \(2^{22} = 4194304\). We define a
function \(xy\), that maps the set of integers \(\{1..2^{22}\}\)
onto the set of couples \(\{0..2^{11}-1\}^{2}\) in lexical order:
\begin{equation}
    xy(i) = (X_{d},Y_{d})
\end{equation}
Next, ordering the couples according to their P' value given in
equation \ref{eq:continuousdiscrete} comes down to the definition
of a permutation p of \(2^{22}\) elements, defined as follows:
\begin{equation}
P'(xy(p(i))) \leq P'(xy(p(j))) \Leftrightarrow i \leq j
\end{equation}
The list \(L'\) of the pairs \( xy(p(i)) \) when \(i\) is running
from \(1\) to \(2^{22}\) is the list of position-ordered couples.
We assign a (non normalized) weight to each element in this list
in the following way:
\begin{equation}
    w_{i} = \textrm{spec}(X_{d}+Y_{d}+1)/(X_{d}+Y_{d}+1)
\end{equation}
with
\begin{equation}
    (X_{d},Y_{d}) = xy(p(i))
\end{equation}
Taking the normalized, cumulative sum in the order of the list
\(L'\), we finally obtain:
\begin{equation}
    r_{i} = \frac{\sum_{j=1}^{i}w_{j}}{\sum_{j=1}^{2^{22}}w_j}
\end{equation}
The new mapping is now defined as:
\begin{equation}
    xy(j) \mapsto [256 r_{p^{-1}(j)}]
\end{equation}
Here we used the inverse permutation \(p^{-1}\), to obtain a final
mapping in lexical order, which is necessary to practically
program the EPROM memory. In this way, each of the bins has very
close to the same summed weight of points in it. We also respect
the order of the value \(P'\) so that a couple with a higher
\(P'\) value can never be assigned to a lower bin than a couple
with a lower \(P'\) value. \emph{Within the family of all possible
lookup tables which respect the order of \(P'\), the proposed
method gives us the optimal solution concerning the flatness of
the efficiency, by construction.}  Indeed, the construction here
proposed makes the efficiencies (the sum of the weights) equal, up
to one single effect, that is: in the order of \(P'\), one has to
decide whether a last point will still be attributed to the lower
bin (and in that case, the extra weight of this single point might
give us a total sum slightly above \(1/2^N\)), or we might decide
to attribute it to the next bin so that the lower bin has a total
sum slightly under \(1/2^N\).  Any lookup table construction
respecting the order of \(P'\) will have to make this decision and
hence at least this error, but in our case, it is the only error.
Moreover, apart from this decision on these pivotal points, the
proposed lookup table is unique.  Hence any other lookup table
will give worse efficiency errors.  Although quite computing
intensive, resolving for the permutation of 4 million pairs of
numbers is now within reach of a modern personal computer and can
be accomplished in a matter of minutes.

Given that we have now the optimal solution concerning efficiency
flatness, we will have to investigate what happened to the
position resolution.  It turns out that the error in position thus
committed is not much larger than the normal quantization error of
the standard method (which is optimal with respect to the position
resolution).  This is understandable, because in the limit of very
high binning in X and Y (for the same resolution in P), that is,
in the continuum limit, the difference between the standard method
and our newly proposed technique vanishes. Indeed, if we work in
the real variables (X,Y), limited to the square \([0,1]^{2}\), let
us define digitized variables:
\begin{equation}
    X_{d} = \left[\frac{X}{\epsilon}\right] ; Y_{d} =
    \left[\frac{Y}{\epsilon}\right] ; \epsilon = \frac{1}{2^{N}}
\end{equation}
with N the very wide word length of the converters. The
accumulated non normalized weight in output channel n (one of the
256), using our new definition of weight, but using the standard
assignment, can be approximated very well by an integral:
\begin{equation}
    \sum_{i\in
    n}
    w_{i}\simeq\int_{0<E<E_{max}}\int_{n-1<256P<n}\textrm{spec}(E)/E
    \frac{dX}{\epsilon}\frac{dY}{\epsilon}
\end{equation}
\begin{equation}
    \sum_{i\in
    n} w_{i}\simeq \frac{1}{256\epsilon^{2}}
    \int \textrm{spec}(E) dE
\end{equation}
We now see that the standard assignment already makes the weights
accumulated in each position bin equal, so we would obtain exactly
the same assignment using our new technique (because the lookup
table doing this is unique), whatever pulse height spectrum is
used. It also means that if we can estimate the loss in resolution
(which will turn out to be insignificant) in a low-bit word
example, this will give us an upper limit on the resolution loss.
We will study a Monte Carlo example to have an idea of what
happens.

\section{Monte Carlo simulation.}

In order to illustrate the problem and its proposed solution, we
simulate the behavior in the case of a low-bit example: we suppose
that the incoming signals X and Y are digitized on 6 bits, and
that we calculate the position on 6 bits (64 positions).  We take
as a spectrum an inverted Raleigh distribution, leading to the
probability density function:
\begin{equation}
    \textrm{spec}(E) = 0.00781 (63-E) e^{-0.0039 (63-E)^{2}}
\end{equation}

\begin{figure}
  \begin{center}
     \includegraphics[width=8.5cm]{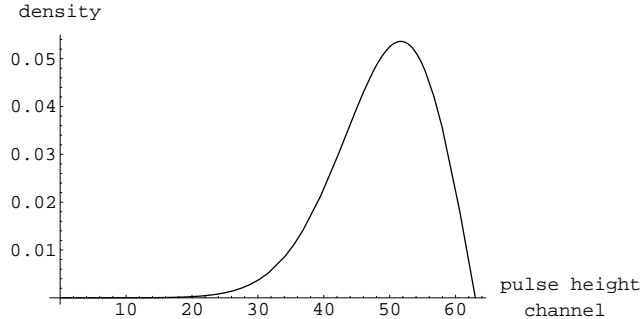}
  \end{center}
  \caption{Artificial distribution that will serve as
  'pulse height spectrum'}
  \label{fig:artificialpulseheightspectrum}
\end{figure}

The precise choice of this distribution doesn't really matter, it
just grossly looks like a true thermal neutron spectrum (using
He-3 as a converter gas and in the proportional region) and is
mathematically well defined. The proposed "spectrum" is displayed
in figure \ref{fig:artificialpulseheightspectrum}. We simulated
\(10^6\) points with a uniform position distribution and an energy
distribution drawn from the above spectrum. Introducing the
digitization according to a 6 bit scale for X and Y, we observe,
using the standard position calculation, a severe digitization
influence (with spikes of the order of 25 \%) on the uniformity of
the response of the system, as displayed in figure 2, full line.

\begin{figure}
  \begin{center}
  \includegraphics[width=8.5cm]{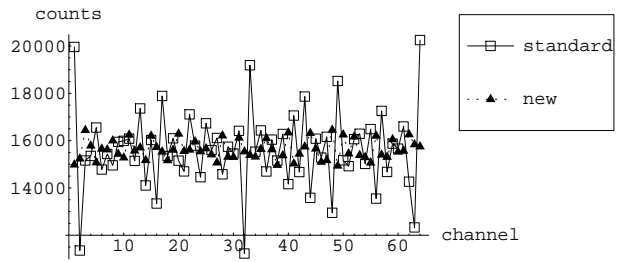}
  \end{center}
  \caption{Result of the standard and new assignment of
  events to position channels.}
  \label{fig:montecarloresult}
\end{figure}

\begin{figure}
  \begin{center}
  \includegraphics[width=8.5cm]{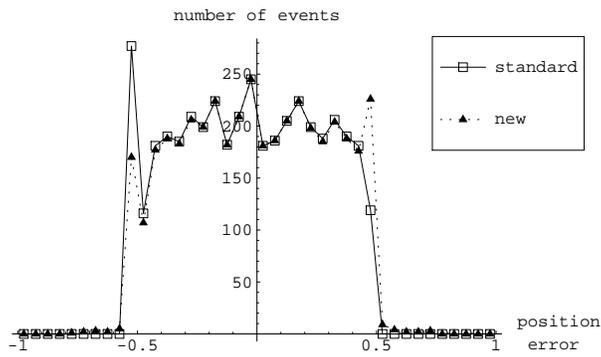}
  \end{center}
  \caption{Distribution of position errors for the standard
  technique and the new technique.}
  \label{fig:montecarloerrors}
\end{figure}

Using the newly proposed technique, we obtain the distribution of
the calculated position in figure \ref{fig:montecarloresult},
dashed line.  The uniformity is clearly superior in the this case.
The price to pay is a very small decrease in resolution. Let us
investigate how severe it is. The quantization error in the case
of the classical calculation is of course close to a uniform
distribution with a standard deviation of \(1/\sqrt{12} =
0.28867\). Using the Monte Carlo data, we find a standard
deviation of 0.292. The standard deviation of the error using the
new mapping is measured to be 0.298 when using a uniform weight
over all couples (X,Y). The two distributions of the position
errors are shown in figure \ref{fig:montecarloerrors}. \emph{We
hence observe a relative loss in resolution of no more than about
2\%.} Essentially, the resolutions can be said to be equivalent,
because this tiny decrease in resolution shouldn't affect the
spatial resolution of the overall system, which shouldn't be
determined by the number of output bins, but by physical processes
and noise limiting the intrinsic resolution of the detector.

If we quantify the non-uniformity by calculating the standard
deviation of the bin contents (using our sample of 1000000 events)
we find the residual standard deviation before quantization is
133.3 counts (per cell) where we expect 125.  Using the standard
way of calculating positions, we have a non-uniformity of 1686.7
counts, and this is reduced to 388.5 counts using the newly
proposed technique.

However, in order for this to hold it is important that the true
spectrum in the data matches closely the spectrum used in the
design of the lookup table.  If we shrink the spectrum of the data
by 5\% (applying a factor 0.95 to the original spectrum used to
construct the table), the good uniformity of the position
histogram is partly gone. The standard technique gives us a
non-uniformity of 1902.3 counts while the new technique obtains
794.2 counts.  A shrinking of 10\% leads to a standard
non-uniformity of 2236.3, while the proposed technique obtains
1424.0 counts.  Although still better, this indicates the need to
match closely the spectrum used in the composition of the table to
the actual spectrum that will be used in order to take the full
advantage of the method. For proportional counters used in thermal
neutron detection, this is a reasonable requirement as the
spectrum is very stable (and also rather broad). On the other
hand, given the very small word lengths in this example, this
simulation is a particularly severe test for our method and in
practice the sensitivity to a change in spectrum is smaller, as
will be shown in the next part.

\section{Experimental verification}

\begin{figure}
  \begin{center}
  \includegraphics[width=8.5cm]{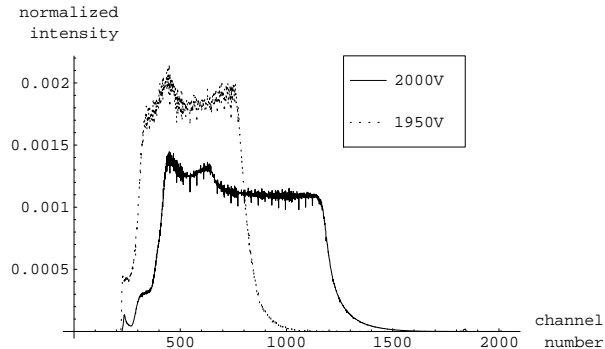}
  \end{center}
  \caption{Normalized Pulse height spectrum of the
  detector at 2000V and at 1950 V.}
  \label{fig:pulseheightmeasure}
\end{figure}

Using an Am-Be thermal neutron source, a 40 cm long
position-sensitive He-3 neutron detector available from
Reuter-Stokes of the type RS-P4-0814-2, and two 3V/pC amplifiers
with an overall gaussian shaping time of 1.4 microseconds, we bias
the detector at 2000V.  This uses almost the full dynamics
available (2048 channels for 5 V) without saturation as
illustrated in figure \ref{fig:pulseheightmeasure}.

\begin{figure}
  \begin{center}
  \includegraphics[width=8.5cm]{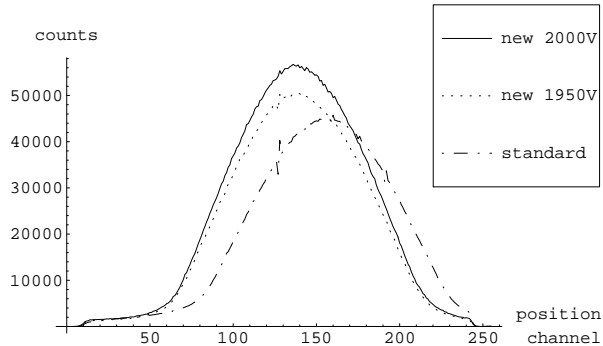}
  \end{center}
  \caption{Images obtained with the standard method and with
    the new method using the  spectrum at 2000 V.  The full curve shows
    the new image at 2000V, and
  the dashed curve shows the new image at 1950 V.  The dash-dotted
  curve gives us the image using the standard method (the number
  of counts has been divided by 20 for this last image in order to
  fit on the figure).  Because of the different manual positioning of the
  neutron source, the standard and new images do not coincide perfectly.}
   \label{fig:standardnewmeasured}
\end{figure}

In order to have very high statistics in a reasonable amount of
time, we put the source very close to the detector; we will have
no uniform illumination, but the smooth bell form image will
indicate local fluctuations in counting efficiency also clearly.
Using the standard algorithm and the new technique, we obtain
different images as displayed in figure
\ref{fig:standardnewmeasured} with the dash-dotted and the full
curve, respectively.

The digitization glitches are clearly visible in the first one and
buried in the statistical noise in the second. (Note that the
position of the source is slightly different, as it was a manual
positioning in between experiments).  Let us now look at the
sensitivity of the spectrum: we change the voltage from 2000 V to
1950 V and use the same lookup table (constructed for use with the
2000 V spectrum). Notice that the applied spectrum (see figure
\ref{fig:pulseheightmeasure}, dotted curve) now has changed by
almost 30\%.  When looking at the resulting image in figure
\ref{fig:standardnewmeasured} (dotted curve), we notice that a
small digitization error appears around channel 128, but it is
still much smaller than if we would have used the standard
algorithm.

We have hence experimentally illustrated that the glitches in
efficiency disappear when applying our new technique.  We've also
demonstrated a certain robustness against differences between the
pulse height spectrum used to construct the lookup table, and the
pulse height spectrum of the detector using the lookup table.


\begin{thebibliography}{10}
\bibitem{articlegeesman}
H. Geesmann et al. Nucl. Instr. Meth. A307 (1991) 413
\bibitem{articletakahashi}
Takahashi et al. Nucl. Instr. Meth. A 373 (1996) 119
\bibitem{articlekoike}
Koike et al. Nucl. Instr. Meth. A272 (1988) 840
\bibitem{articleberliner}
Berliner et al. Nucl. Instr. Meth. 184 (1981) 477
\bibitem{articlemori}
C. Mori et al. Nucl. Instr. Meth. 299 (1990) 128
\bibitem{articleuritani}
A. Uritani et al. Nucl. Instr. Meth. 353 (1994) 246
\end{thebibliography}
\end{document}